\documentclass[onecolumn,showpacs,preprintnumbers,superscriptaddress,amsmath,amssymb]{revtex4}
\usepackage[dvips]{graphicx}
\usepackage{latexsym,amssymb,amsmath,epsfig,bm,times,psfrag}
\usepackage{color}
\usepackage[bookmarksnumbered,bookmarksopen,colorlinks,citecolor=red,linkcolor=blue]{hyperref}
\usepackage{epstopdf}
\usepackage[capitalize]{cleveref}
\newcommand{\Fig}[1]{Fig.~\ref{#1}}
\newcommand{\eq}[1]{Eq.~(\ref{#1})}
\newcommand{\sect}[1]{Sec.~\ref{#1}}

\renewcommand{\part}{{\rm part}}
\renewcommand{\vec}{\boldsymbol}
\newcommand{\be}{\begin{equation}}
\newcommand{\ee}{\end{equation}}
\newcommand{\bear}{\begin{eqnarray}}
\newcommand{\eear}{\end{eqnarray}}
\newcommand{\ba}{\begin{array}}
\newcommand{\ea}{\end{array}}

\begin{document}

%\preprint{00-000}

\title{Event-shaped-dependent cumulants in p-Pb collisions at 5.02 TeV}

\author{De-Xian Wei}
\email{dexianwei@gxust.edu.cn}
\affiliation{School of Science, Guangxi University of Science and Technology, Liuzhou, 545006, China}
\author{Li-Juan Zhou}
\email{zhoulijuan05@hotmail.com}
\affiliation{School of Science, Guangxi University of Science and Technology, Liuzhou, 545006, China}
\date{\today}% It is always \today, today, but any date may be explicitly specified

\begin{abstract}
\noindent
In this paper, we present a novel event-shaped cumulants (ESC) response approach,
based on a multi-phase transport (AMPT) model simulations, to analyze p-Pb collisions at $\sqrt{s_{NN}}$= 5.02 TeV.
We find that the Pearson coefficients between the subset cumulants of the final harmonics $v_{2}\{2k\}~(k=1,2,3,4)$
and the subset cumulants of initial eccentricity $\varepsilon_{2}\{2k\}$ in the ESC basis are significantly enhanced.
These Pearson coefficients are strongly-dependent on the charged multiplicity, the set number of events (SNE),
and only weakly-dependent on the order of the multi-particle cumulants (two-particles, four-particles, and so on).
Our results show that the ESC method can suppress the event-by-event fluctuations.

\par
\par

\end{abstract}

% PACS, the Physics and Astronomy Classification Scheme.

\maketitle

%=======================================Document Begin=========================================

%======================================= Introduction==========================================
\section{Introduction}
\label{sec:sec1}
One of the most remarkable achievements in ultra-relativistic heavy-ion experiments is
the production of a fluidlike quark-gluon system, usually referred to as the quark-gluon
plasma (QGP)~\cite{Adams:2004pdo,Song:2011acs}. Collective harmonics flow plays a major role in probing the properties of
QGP at the Relativistic Heavy Ion Collider (RHIC) of the Brookhaven National Laboratory
(BNL)~\cite{Abelev:2009lrr}, and at the Large Hadron Collider (LHC) of the European
Organization for Nuclear Research (CERN)~\cite{ATLAS:2012mot}. These collective flows may
be well described by the hydrodynamic model~\cite{Shen:2015saf,Gale:2013eaf,McDonald:2017hpf,Calzetta:2020dtt,Monnai:2021qeo},
which show that the collective flow behavior is very sensitive to the energy density,
and its fluctuations of the initial state.

To study the fluctuations of the initial state, Glauber, various hydrodynamic and transport models
have been suggested, as well as approximate response relations between the final harmonics
flow $v_{n}$ and the initial eccentricity $\varepsilon_{n}$ of large
systems~\cite{Enterria:2020pit,Alver:2010tfi,Noronha-Hostler:2015dbi,Yan:2017ivm,De:2018hri,Bozek:2016tmc,Floerchinger:2014soi,Roch:2021foa}.
The hydrodynamic expectation is that $v_{n}/\varepsilon_{n}$ ratio increases monotonically with the transverse density across different collision energies and systems, and a violation of such a scaling may indicate an incorrect modeling of the initial transverse area and/or the
azimuthal anisotropies  $\varepsilon_{n}$~\cite{Enterria:2020pit}.

Linear response relations for small systems, e.g., p-Pb and peripheral Pb-Pb collisions, have
been also discussed in our previous work~\cite{De:2020rli}, and it has also been pointed out
that a linear relation between the global $v_{2}\{2k\}~(k=1,2)$
and the global $\varepsilon_{2}\{2k\}$ in large hydrodynamic systems~\cite{Qiu:2012hea},
agrees with ALICE results (with the Kharzeev-Levin-Nardi initial conditions)~\cite{Aamodt:2011hha}.
Furthermore, studies of multi-particle cumulants in the p-p~\cite{Khachatryan:2016efc}, p-Pb~\cite{Khachatryan:2015efcm}
and p/d-Au~\cite{Aidala:2018mom} systems have revealed collective behavior similar
to that found in heavier systems~\cite{Chatrchyan:2013moh}.
The hydrodynamic~\cite{Yan:2013ufe} predictions for the multi-particle cumulants in
p-Pb collisions, especially, the $v_{2}\{4\}/v_{2}\{2\}$ ratio is a fact that suggests a direct
correlation of the final state harmonics with the initial state eccentricity.

The Pearson coefficient is a suitable quantity to describe such fluctuating response relation
between the final harmonics flow $v_{n}$ and the initial eccentricity $\varepsilon_{n}$.
Studies of nonzero Pearson coefficient in large systems have been presented ~\cite{Niemi:2013edo,Liu:2019pca}, and also in high multiplicity p-Pb systems~\cite{Yan:2014aad}.
However, the observed event-by-event Pearson coefficient is negligible in peripheral Pb-Pb collisions, which themselves include low multiplicity and may be referred to as a small system~\cite{Rao:2019bpo}.

In fact, it has been argued that if the system size is too small, and the lifetime
is too short, a hydrodynamic approach with local isotropization~\cite{Schenke:2017ooc}
may be not suitable. In addition, the presence of a collective flow in such a small
systems is still being debated~\cite{Dusling:2016ncp}. Indeed, there are tremendous
fluctuations in such small systems due to low multiplicity distribution, and
a question naturally arises on whether a fluidlike QGP may be actually
created ~\cite{Nagle:2018ssc}.

The nature of the fluctuating response between the final collective harmonics $v_{n}$
and the initial eccentricity $\varepsilon_{n}$ for small systems is
still unclear both in hydrodynamic and transport models. On the other hand, the analysis of hydrodynamic models has revealed that even thought the system is small at the beginning,
it may expand into a larger system in the final-states with large energy and entropy~\cite{Heinz:2019hfi}. Of course, one must be careful in assessing the flow fluctuations versus
the nonflow ones in these small systems.
One proposed solution to resolve the above questions about the smallest droplet of the QGP is to run polarized ion beams~\cite{Bozek:2018efi}, which is certainly an appealing suggestion but may take years to create the proper infrastructure to do so.
To know what happened in such polarized ion collisions, one need to solve the dynamics with spin. The hydrodynamic model with spin has been calculated in the large systems~\cite{Saha:2020fav,Bhadury:2021ndi}, while it still needs to be carefully verified in the small systems.

If we assume that a collective flow exists in these small reaction systems, then the analysis of
fluctuations represents a very important tool to study the properties of the medium.
To this aim, we here present a new type of event shaped cumulants (ESC) response approach for
p-Pb collisions. This work follows our previous work, where we have discussed the response
relations in p-Pb systems with event-by-event simulations~\cite{De:2020rli}.
This work mainly discusses the event-by-event initial eccentricity fluctuations
(except for the multiplicity fluctuations in~Fig.~\ref{fig2}), and does not intend to
discuss the effects caused by other sources of fluctuations. In event-by-event small
collisions, the  fluctuations of the collective flow are mostly originating from the
fluctuations of the initial geometry fluctuations and of the density. Therefore, we focus
on the subset of events (SOE) cumulants response in p-Pb collisions using an ESC analysis.
In the following, the ESC is proposed as an effective observation to suppress
event-by-event initial geometry fluctuations.

The paper is organized as follows: In \sect{sec:sec2} we briefly describe
the ESC response theory in a multi-phase transport (AMPT)
model~\cite{Lin:2004amt}, which is then used in the simulations.
The numerical results about the correlations between subsets of $v_{2}\{2k\}~(k=1,2,3,4)$
and subsets of $\varepsilon_{2}\{2k\}$ are presented in \sect{sec:sec3}.
All of the produced charged pions in the calculations are chosen with
0.3 $< p_{T} <$ 3.0 GeV, $|\eta|<$1.0 and $80\leq M\leq 120$
(except for the multiplicity fluctuations in~Fig.~\ref{fig2}). Finally, we summarize our
main results in \sect{sec:sum}.

%=================================== theory model =========================================
\section{Materials and Methods}
\label{sec:sec2}
The AMPT model is a hybrid transport model for high-energy heavy ion collisions~\cite{Lin:2004amt}.
The AMPT model can be produced particles in difference stages, from initial to final states.
In AMPT model, the initial state particle distributions are generated by HIJING model~\cite{Wang:1991hac}. For this study, string
melting is considered so that the produced hadrons from HIJING model are further converted
into valence quarks and anti-quarks. Right before parton scatterings, we record the generated
energy density profile of the system $\rho(\vec x,\tau_0)$, as the initial state of medium
evolution. Initial state eccentricities of each event are then calculated with respect
to \eq{eq:en}. Parton scatterings, and accordingly the space-time evolution of
QGP, are determined via ZPC parton cascade model~\cite{Zhang:1998zac}.
After that, quarks and anti-quarks combine to form hadrons via a spatial coalescence model when scatterings stop.
Hadronic phase of the system evolves according to a relativistic transport model until hadrons freeze-out.
Note that it not only includes the medium matter but also includes the nonflow effects which are produced in hard scattering and hadron decay.
To study the fluctuations of hydrodynamic property, we need a good physical quantity to describe the collective property of the medium,
one is the fluctuating harmonic flow.
To theoretically investigate these fluctuating behaviors, AMPT is an appropriate tool for the collective response analysis.
More details about the AMPT model can be found in Ref.~\cite{Lin:2004amt}.

To study the event-by-event fluctuations, we start from the estimators of the harmonic flow $V_n$
and the initial eccentricity $E_{n}$, which have been studied in Ref.~\cite{Alver:2010gr},
\be
\label{eq:vn}
V_n = v_n e^{in\Psi_n} \equiv \int \frac{d\phi}{2\pi} e^{in\phi_p} f(\phi_p)\,,
\ee
and Ref.~\cite{Teaney:2010vd},
\be
\label{eq:en}
E_n=\varepsilon_n e^{in\Phi_n}\equiv -{\int d^2\vec x_\perp \rho(\vec x_\perp, \tau_0) r^n e^{in\phi}
\over
\int d^2\vec x_\perp r^n\rho(\vec x_\perp, \tau_0)}\,~~(n\geq2).
\ee
where the magnitudes $v_n$ and $\varepsilon_n$ fluctuate on an event-by-event basis.
The magnitudes $v_n$ and $\varepsilon_n$ in \eq{eq:vn} and \eq{eq:en} can be accurately calculated by the event plane method~\cite{Ma:2021pce}, respectively.
In this work, we focus on the second-order harmonic for the analyses.
In turn, $V_{2}$ is found to be very sensitive to the event-by-event fluctuating initial eccentricity~\cite{De:2018hri,Nie:2019ioi},
e.g., $v_{n}\{EP\}= \langle cos[n(\varphi-\Psi_n)]\rangle$
and $\varepsilon_{n}= \sqrt{\langle r^{n}cos(n\phi) \rangle^{2}+\langle r^{n}sin(n\phi) \rangle^{2}}/\langle r^{n} \rangle$,
where $\varphi$ is the azimuthal angle of the produced particle, and $\Psi_n$ is the corresponding symmetry event plane angle.

\begin{figure}[h]
\includegraphics[width=0.40\textwidth]{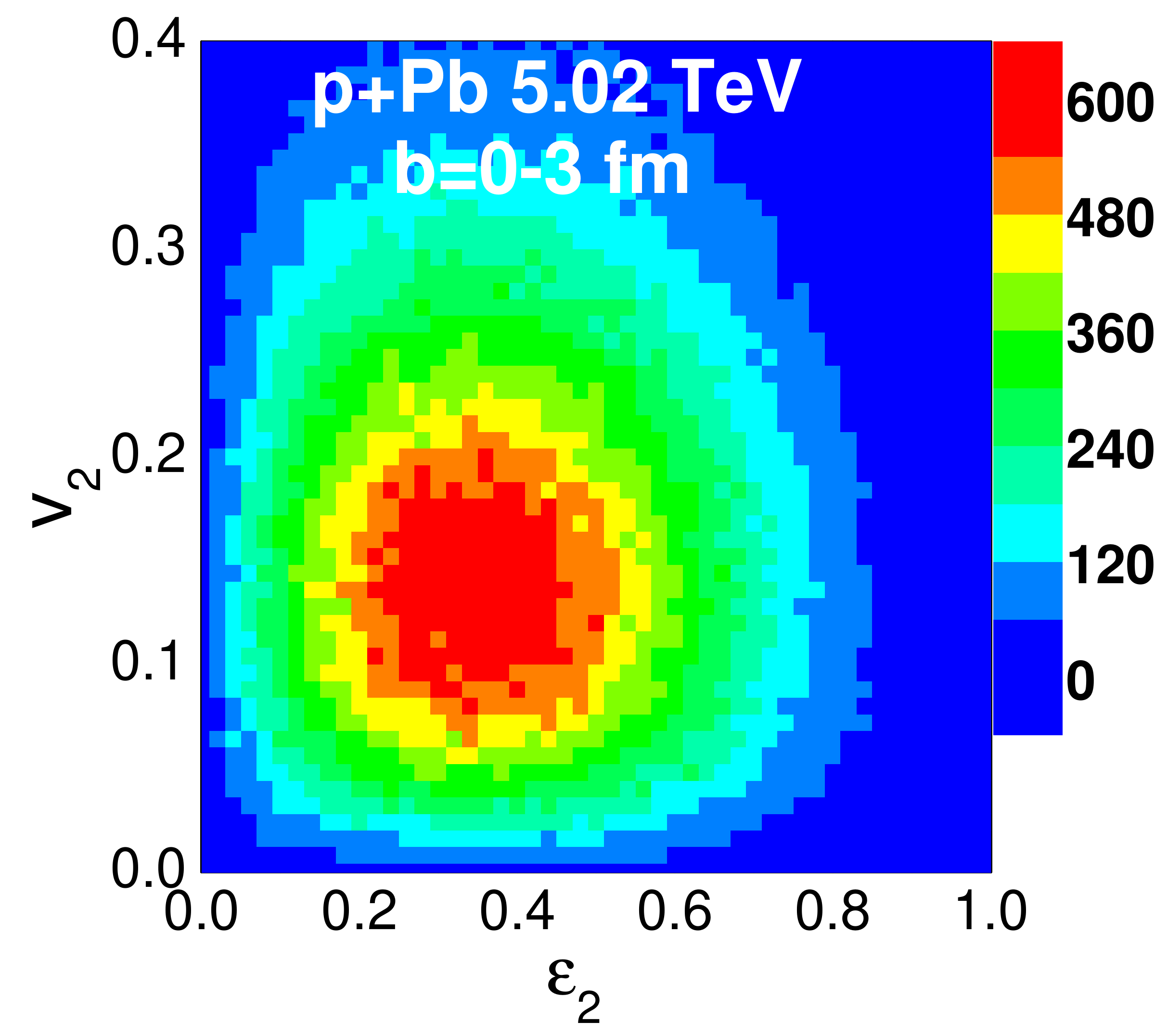}
\caption{(Color online)
Scatter plot of event-by-event $v_2$ from AMPT simulations for p-Pb (b=0-3 fm) collisions, as a function of $\varepsilon_2$.
}
\label{fig1}
\end{figure}

Let us first explore the behavior of fluctuations in small systems using \eq{eq:vn} and \eq{eq:en}.
Fig.~\ref{fig1} shows a scatter plot of event-by-event $v_2$ from AMPT simulations for p-Pb (b=0-3 fm) collisions, as a function of $\varepsilon_2$.
One can see a plump distribution of points, which means that the event-by-event final harmonic
flow is not correlated to the initial eccentricity, i.e. giant fluctuations and/or nonflow effects cannot
be ignored in analyzing the phenomenon. The corresponding Pearson coefficient, evaluated as
in Ref.~\cite{De:2018hri} is rather low $C(\varepsilon, v)\approx 0.04$. Similar results
have been discussed in Ref.~\cite{Rao:2019bpo}, using a hydrodynamic model. Even
though hydrodynamic arguments suggest that a cubic response should be considered, due to
the initial fluctuations~\cite{Rao:2019bpo}, a complete understanding of the
type of the response in such small system is still missing. In turn, such initial
fluctuations may be described by the phase of $V_{2}$, such that the event plane
may be the complex plane.

Another point to be taken account is the fact that the physics of p-A collisions
may actually lie in the transition region between pure hydrodynamics and
kinetic-particle evolution, as suggested by a kinetic theory analysis~\cite{Kurkela:2019fia}.
The hydrodynamic model indeed fails to describe these kinetic-particle effects,
which are referred to as nonflow effects. In turn, a weak Pearson coefficient
implies that nonflow effects cannot be ignored in  small systems. We conclude that
the event-by-event Pearson coefficient may not represent the proper quantity to
describe collective behaviors in such small systems, due to the finite multiplicity
statistics.

To study the effective Pearson coefficient, an ESC method
may be introduced. The multi-particle cumulants method is a useful tool in
studying the global properties of correlations along the azimuthal direction
in high-energy collisions~\cite{Aaboud:2017mom}.

Let us start with the calculation of $2k$-particle (where $k=1,2,3,4$) azimuthal
 correlations~\cite{Borghini:2001nmf,Bilandzic:2011faw,Aaboud:2017mom}
\begin{eqnarray}\label{eq:se}
\langle 2 \rangle &\equiv& \langle e^{i2(\phi_{1}-\phi_{2})}\rangle = \frac{|Q_{n}|^{2}-M}{M(M-1)},  \nonumber \\
\langle 4 \rangle &\equiv& \langle e^{i2(\phi_{1}+\phi_{2}-\phi_{3}-\phi_{4})}\rangle = \frac{|Q_{n}|^{4}+|Q_{2n}|^{2}-2\cdot Re[Q_{2n}Q_{n}^{*}Q_{n}^{*}]-4(M-2)|Q_{n}|^{2}-2M(M-3)}{M(M-1)(M-2)(M-3)},  \nonumber \\
\langle 6 \rangle &\equiv& \langle e^{i2(\phi_{1}+\phi_{2}+\phi_{3}-\phi_{4}-\phi_{5}-\phi_{6})}\rangle \nonumber \\
&=& \{|Q_{n}|^{6}+9|Q_{2n}|^{2}|Q_{n}|^{2}-6\cdot Re[Q_{2n}Q_{n}Q_{n}^{*}Q_{n}^{*}Q_{n}^{*}]+4[Re[Q_{3n}Q_{n}^{*}Q_{n}^{*}Q_{n}^{*}]  \nonumber \\
&-& 3\cdot Re[Q_{3n}Q_{2n}^{*}Q_{n}^{*}]] + 2[9(M-4)\cdot Re[Q_{2n}Q_{n}^{*}Q_{n}^{*}]+2|Q_{3n}|^{2}] - 9[|Q_{n}|^{4}+|Q_{2n}|^{2}](M-4) \nonumber \\
&+& 18[|Q_{n}|^{2}(M-2)(M-5)]-6M(M-4)(M-5)\} \nonumber \\
&/& \{M(M-1)(M-2)(M-3)(M-4)(M-5)\},  \nonumber \\
\langle 8 \rangle  &\equiv& \langle e^{i2(\phi_{1}+\phi_{2}+\phi_{3}+\phi_{4}-\phi_{5}-\phi_{6}-\phi_{7}-\phi_{8})}\rangle \nonumber \\
&=& \{ |Q_{n}|^{8}-12\cdot Q_{2n}Q_{n}Q_{n}Q_{n}^{*}Q_{n}^{*}Q_{n}^{*}Q_{n}^{*} + 6\cdot Q_{2n}Q_{2n}Q_{n}^{*}Q_{n}^{*}Q_{n}^{*}Q_{n}^{*}  \nonumber \\
&+& 16\cdot Q_{3n}Q_{n}Q_{n}^{*}Q_{n}^{*}Q_{n}^{*}Q_{n}^{*} - 96\cdot Q_{3n}Q_{n}Q_{2n}^{*}Q_{n}^{*}Q_{n}^{*} -12\cdot Q_{4n}Q_{n}^{*}Q_{n}^{*}Q_{n}^{*}Q_{n}^{*}  \nonumber \\
&-& 36\cdot Q_{2n}Q_{2n}Q_{2n}^{*}Q_{n}^{*}Q_{n}^{*} + 96(M-6)\cdot Q_{2n}Q_{n}Q_{n}^{*}Q_{n}^{*}Q_{n}^{*} +72\cdot Q_{4n}Q_{2n}^{*}Q_{n}^{*}Q_{n}^{*} \nonumber \\
&+& 48\cdot Q_{3n}Q_{n}Q_{2n}^{*}Q_{2n}^{*} - 64(M-6)\cdot Q_{3n}Q_{n}^{*}Q_{n}^{*}Q_{n}^{*} + 192(M-6)\cdot Q_{3n}Q_{2n}^{*}Q_{n}^{*} -96\cdot Q_{4n}Q_{3n}^{*}Q_{n}^{*}  \nonumber \\
&-& 36\cdot Q_{4n}Q_{2n}^{*}Q_{2n}^{*} - 144(M-7)(M-4) Q_{2n}Q_{n}^{*}Q_{n}^{*} + 36 |Q_{4n}|^{2}+64|Q_{3n}|^{2}|Q_{n}|^{2}  \nonumber \\
&-& 64(M-6) |Q_{3n}|^{2} + 9 |Q_{2n}|^{4}+36|Q_{n}|^{4}|Q_{2n}|^{2}-144(M-6)|Q_{2n}|^{2}|Q_{n}|^{2}  \nonumber \\
&+& 72(M-7)(M-4)(|Q_{2n}|^{2} + |Q_{n}|^{4})-16(M-6)|Q_{n}|^{6}   \nonumber \\
&-& 96(M-7)(M-6)(M-2)|Q_{n}|^{2}+24M(M-7)(M-6)(M-5) \} \nonumber \\
&/& \{M(M-1)(M-2)(M-3)(M-4)(M-5)(M-6)(M-7)\}.
\end{eqnarray}
Here, $n$=2, and $\phi_{j}$ is the azimuthal angle of the $j$-th particle transverse momentum and pseudorapidity region in a single event, and the $Q$-vector is $Q_{n}=\sum_{j}^{M}e^{in\phi_{j}}$.
$M$ is the number of particles.
The single bracket indicates that all of present particles are averaged out in a single event.
Using the calculated multi-particle azimuthal correlations, the second-order
cumulants $c_{2}\{2k\}~(k=1,2,3,4)$~\cite{Borghini:2001nmf,Bilandzic:2011faw,Aaboud:2017mom}
are obtained after subtracting the correlations between $2k~(k \geq 1)$ particles and averaged in an SOE, i.e.,
\begin{eqnarray}\label{eq:ce}
c_{2}\{2\}(i) &=& \langle\langle 2 \rangle\rangle_{i}, \nonumber \\
c_{2}\{4\}(i) &=& \langle\langle 4 \rangle\rangle_{i}-2\cdot\langle\langle 2 \rangle\rangle_{i}^{2}, \nonumber \\
c_{2}\{6\}(i) &=& \langle\langle 6 \rangle\rangle_{i}-9\cdot\langle\langle 2 \rangle\rangle_{i}\langle\langle 4 \rangle\rangle_{i}+12\cdot\langle\langle 2 \rangle\rangle_{i}^{3}, \nonumber \\
c_{2}\{8\}(i) &=& \langle\langle 8 \rangle\rangle_{i}-16\cdot \langle\langle 6 \rangle\rangle_{i}\langle\langle 2 \rangle\rangle_{i} - 18\cdot\langle\langle 4 \rangle\rangle_{i}^{2} +144\cdot \langle\langle 4 \rangle\rangle_{i}\langle\langle 2 \rangle\rangle_{i}^{2}-144\cdot\langle\langle 2 \rangle\rangle_{i}^{4}.
\end{eqnarray}
where $i$ is the $i$th SOE cumulant. The double brackets mean that the event-by-event
correlations are averaged in the $i$th SOE cumulant.

In this work, we focus on the multi-particle cumulants in ESC analysis, rather
than in total events. Calculations refer to a subset of total events and, as a result,
the total calculations are divided in multiple ESC. In our simulations,
we thus set the number of events (denoted by SNE) included in a SOE cumulant
(eg. 20,~50,~200,~500 or more/less events per SOE cumulant).
For a fixed number of total events, we have different numbers of SOE for different
SNE parameters, e.g., if the total number of events is  5$\times 10^{5}$ and
SOE is set to 25000, 10000, 2500 and 1000, then we have SNE=20, 50, 200 and 500, respectively.
If the SOE cumulants $c_{2}\{2k\}(i)~(k=1,2,3,4)$ are free of nonflow correlations
and only weakly affected by number fluctuations,
they can be used to estimate the SOE harmonics $v_{2}$, which
can be written as
\begin{eqnarray}\label{eq:localcumulants}
v_{2}\{2\}(i) &=& \sqrt{c_{n}\{2\}(i)}, \nonumber \\
v_{2}\{4\}(i) &=& \sqrt[4]{-c_{n}\{4\}(i)}, \nonumber \\
v_{2}\{6\}(i) &=& \sqrt[6]{c_{n}\{6\}(i)/4}, \nonumber \\
v_{2}\{8\}(i) &=& \sqrt[8]{-c_{n}\{8\}(i)/33}.
\end{eqnarray}
Here, the collective flow is estimated using a cumulants expansion of multi-particle correlations,
without determining the orientation of the event plane. The reason for this is that if
the particles are correlated with the orientation of the event plane,
then the information about the event plane orientation is erased.
Note that the sample in the evaluation of SOE cumulants is not randomly selected.
Events are sorted according to the eccentricity, from minimum to maximum,
after the evolution of event-by-event, and then the event subsets are divided.
Finally, the SOE cumulants harmonics flow $v_{2}\{2k\}(i)$ are used for the ESC analysis.
The calculation of ESC for the initial eccentricity is similar to that of the ESC for final harmonics, however the initial partons are used.

In order to estimate the correlation between two different samples, the Pearson
coefficient $C(A, B)$ is defined in ESC basis, as
\begin{widetext}
\begin{equation} \label{eq:pearsonc}
C(A, B)=\frac{\sum_{i=1}^{N}[A(i)-\langle A \rangle][B(i)-\langle B \rangle]} {\sqrt{\sum_{i=1}^{N}[A(i)-\langle A \rangle]^{2}} \sqrt{\sum_{i=1}^{N}[B(i)-\langle B \rangle]^{2}}}.
\end{equation}
\end{widetext}
where $i$ is the $i$-th sample of SOE and $\langle\ldots \rangle$ denotes
the total SOE events averaged. $N$ is the total number of SOEs.
The variables $A$ and $B$ could be samples such as the SOE cumulants
of initial eccentricity or the SOE cumulants of final harmonics.
Notice that the cumulants are real numbers, and thus the Pearson coefficient
in \eq{eq:pearsonc} is a real number.
In this case, $C(A, B)=1$ means that the two samples are linearly fully correlated,
whereas and $C(A, B)=0$ corresponds to completely linearly uncorrelated quantities.
In this work, we only study the degree of linear correlation of the samples, without
entering the discussing about the possible nonlinear correlations, which are still under debate
in small systems~\cite{Rao:2019bpo,De:2020rli}.

\begin{figure*}%[h]
\begin{center}
\includegraphics[width=0.92\textwidth]{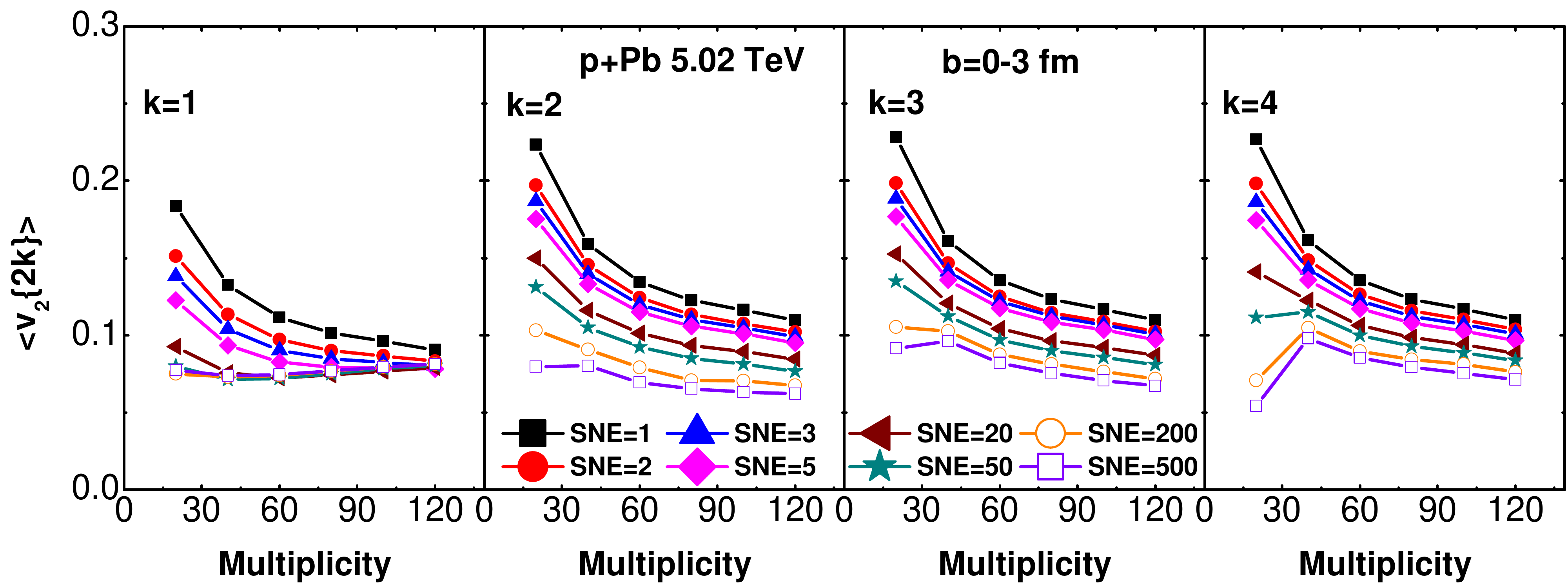}\\
\includegraphics[width=0.92\textwidth]{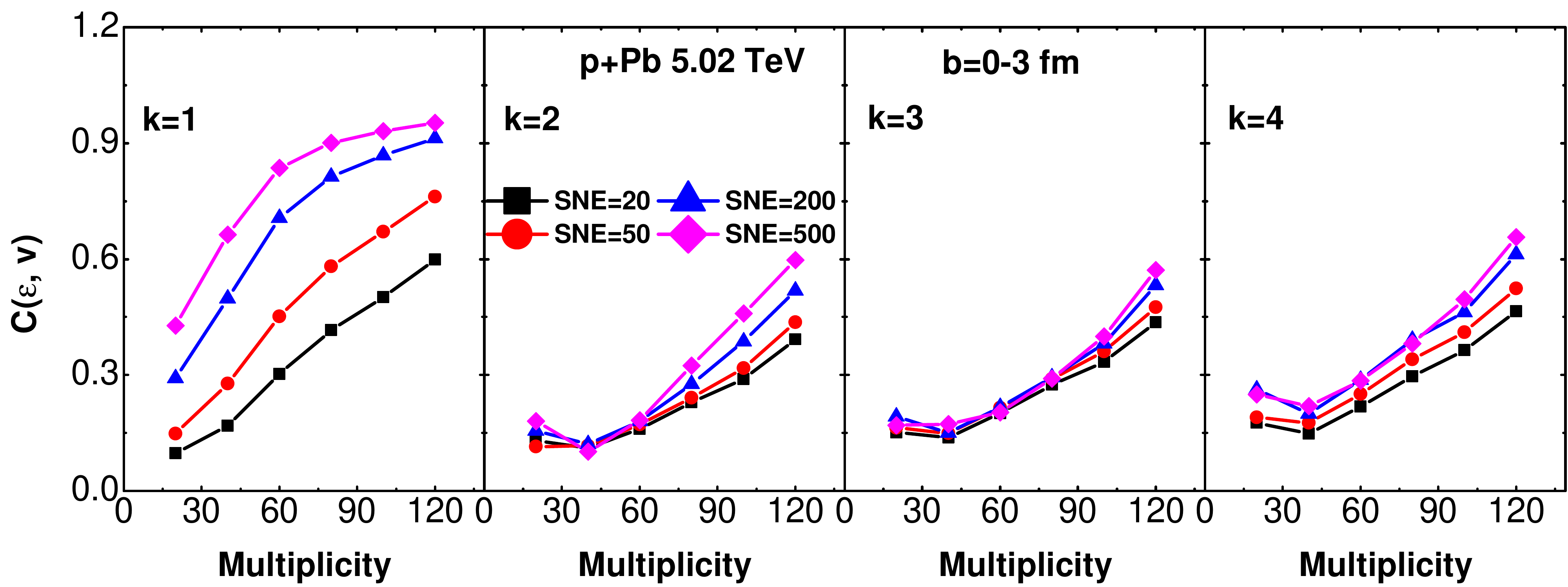}
\caption{(Color online)
Cumulants flow $\langle v_{2}\{2k\}\rangle$ and Pearson coefficients $C(\varepsilon, v)$ as functions of the charge multiplicity in p-Pb (b=0-3 fm, fixed number of
events) collisions, respectively. Up panels: for the results of cumulants flow. Down panels: for the results of Pearson coefficients.
}
\label{fig2}
\end{center}
\end{figure*}

In this work, the Pearson coefficients between the SOE cumulants of the final harmonics and
the SOE cumulants of the initial eccentricity are investigated for p-Pb (b=0-3 fm)
collisions at $\sqrt{s_{NN}}$ = 5.02 TeV. We analyze the output charged
pions using AMPT model~\cite{Lin:2004amt}. It takes
the specific shear viscosity $\eta/s=0.273$, which is calculated by the Lund string fragmentation parameters in AMPT,
i.e., $a=0.5$, $b=0.9$ GeV$^{-2}$, $\alpha_{s}$ =0.33 and $\mu$ = 3.2 fm$^{-1}$,
which are suited for LHC~\cite{De:2018hri,De:2020rli}.
Here, we use different SNE parameters to simulate data and obtain
the cumulants flow $\langle v_{2}\{2k\}\rangle$ and Pearson coefficients $C(\varepsilon, v)$ in p-Pb (b=0-3 fm) collisions.
The impact parameters are controlled by the transverse distance of the overlap in the initial collisions space, and do not depend
on the participant multiplicity $M$. As a consequence, our analysis includes
a wide multiplicity range, from low $M$ to large ones,
with a predominant range $M\leq 120$.  The cumulants flow $\langle v_{2}\{2k\}\rangle$ are calculated from \eq{eq:localcumulants}. And the Pearson coefficients $C(\varepsilon, v)$ between subset of $v_{2}$ and subset of $\varepsilon_{2}$ are calculated from \eq{eq:pearsonc}. Both the cumulants flow $\langle v_{2}\{2k\}\rangle$ and Pearson coefficients $C(\varepsilon, v)$ dependent on the charged multiplicity, as shown in \Fig{fig2}.
They are strongly-dependent on the set number of events (SNE),
and only weakly-dependent on the order of the multi-particle cumulants (two-particles, four-particles, and so on).
The $\langle v_{2}\{2k\}\rangle$ decrease with the charged multiplicity (except for lower multiplicity $M<40$), while $C(\varepsilon, v)$ increase with the charged multiplicity.
These results within smaller multiplicity are similar in Ref.~\cite{Jia:2017rlm}, due to the residual
nonflow effects and/or fluctuations. Such newly developed subevent methods~\cite{Jia:2017rlm} is failed to suppress the residual nonflow contributions in small collisions.
Note that the subevent method in Ref.~\cite{Jia:2017rlm} are ranged all sub-samples in a single event,
while our results of sub-samples are acting at different events.
In this work, we focus on suppressions of fluctuations with the subset of standard cumulants in the analysis.
Therefore, for each chosen samples, its charged multiplicity is ranged in $M_{ch}\in [80, 120]$. Such settings are used in~\cref{fig3,fig4,fig5,fig6}.
Note also that in this paper, we do not attempt to compare simulations with experimental data,
but rather to explore how the correlations between the subset of final harmonic and subset of initial eccentricity are influenced by the initial fluctuations.

%=======================================calculations and analysis=========================================
\section{Results}
\label{sec:sec3}

\begin{figure*}[t]
\begin{center}
\includegraphics[width=0.860\textwidth]{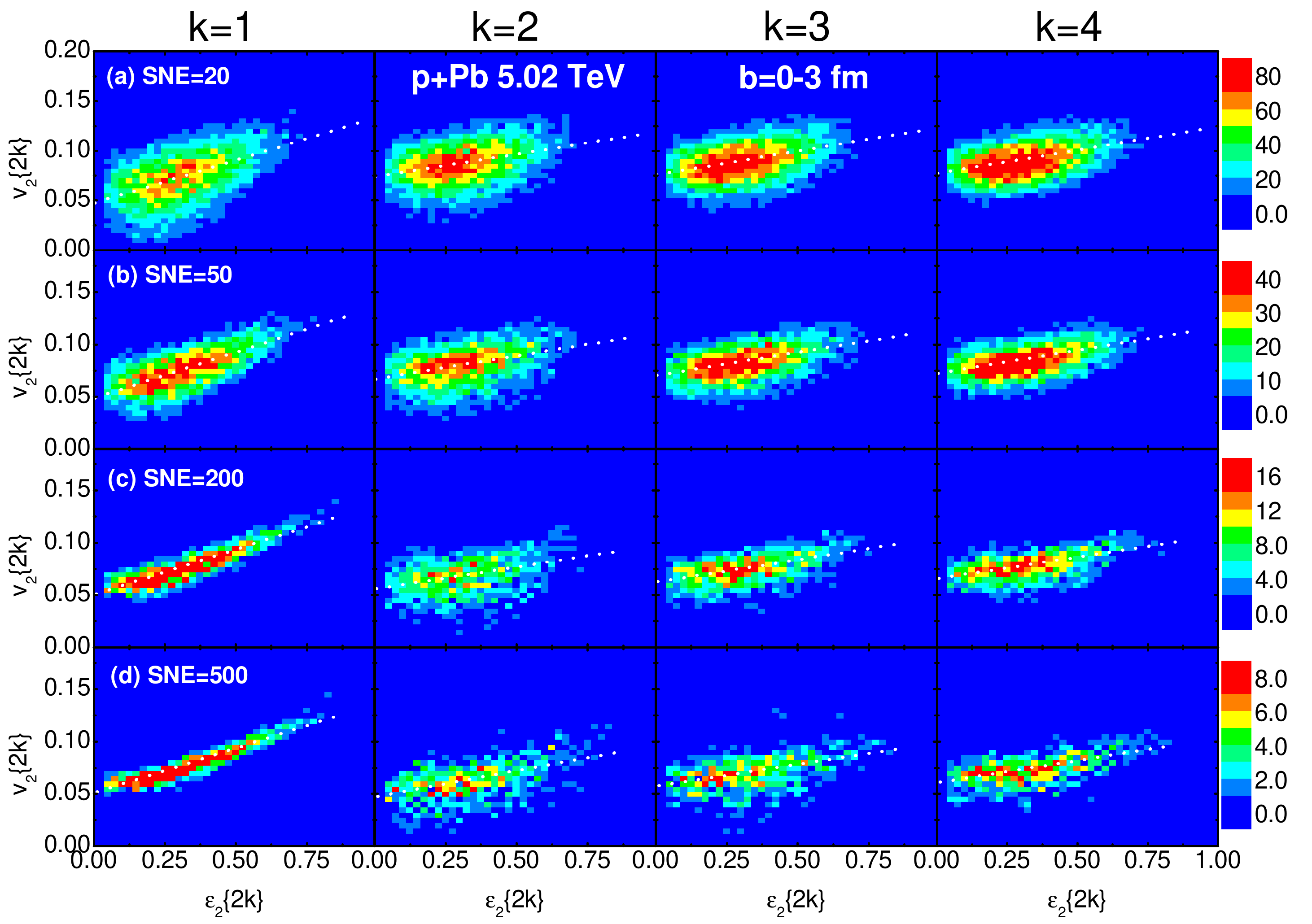}
\caption{(Color online)
Scatter plot of SOE cumulants $v_{2}\{2k\}~(k=1,2,3,4)$ from AMPT simulations for p-Pb (b=0-3 fm) collisions (at fixed number of events),
as a function of SOE cumulants $\varepsilon_{2}\{2k\}$.
The numbers $k=$1,~2,~3,~ and 4 correspond to the two-particle, four-particle, six-particle and eight-particle cumulants, respectively.
The number of SNE is set to 20, 50, 200, and 500, respectively.
}
\label{fig3}
\end{center}
\end{figure*}

\begin{figure*}%[tp]
\begin{center}
\includegraphics[width=0.860\textwidth]{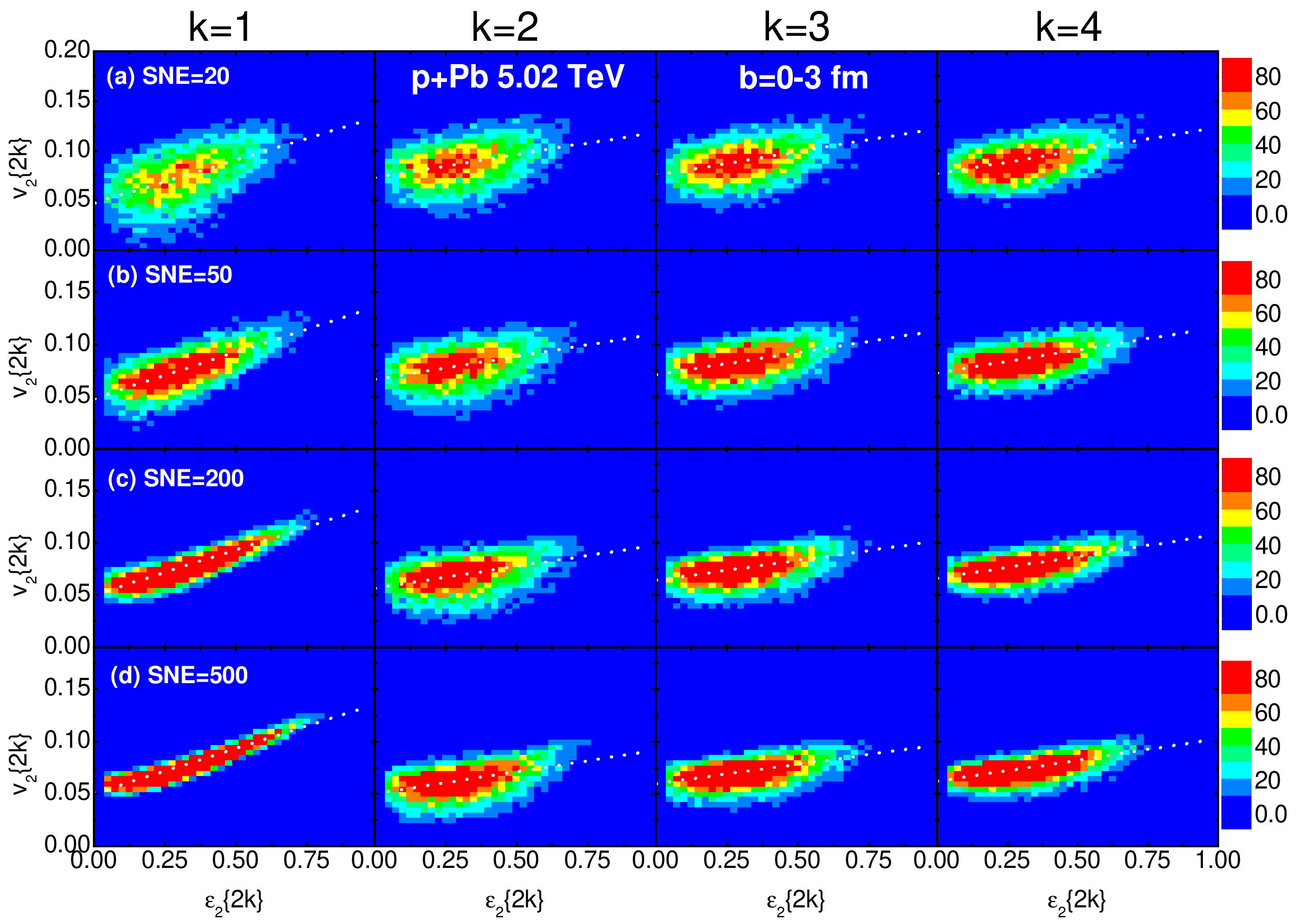}
\caption{(Color online)
As in \Fig{fig2}, but for fixed scatter points (for different SNE, the number of scatter
points is set by consistency).
}
\label{fig4}
\end{center}
\end{figure*}

The two-dimensional histograms in Figs.~\ref{fig3} and \ref{fig4} show the correlation between the subset of $v_{2}\{2k\}~(k=1,2,3,4)$
and the subset of $\varepsilon_{2}\{2k\}$ for p-Pb (b=0-3 fm) collisions, respectively.
To illustrate the nature of these correlations, we report the scatter plot for four different cases:
(a) SNE=20; (b) SNE=50; (c) SNE=200 and (d) SNE=500.

Figure~\ref{fig3} shows the scatter plot of subset of $v_{2}\{2k\}~(k=1,2,3,4)$ from ESC AMPT simulations for p-Pb (b=0-3 fm) collisions,
as a function of subset of $\varepsilon_{2}\{2k\}$.
The numbers $k=$1, 2, 3, and 4  correspond to two-particle, four-particle, six-particle and eight-particle cumulants, respectively.
The SNE is equal to 20, 50, 200 and 500 single events, respectively.
Each point in \Fig{fig3} corresponds to a SOE cumulant.
The white dashed line is the result of a linear least-square fitting. The value of the linear response coefficients is about 0.08.
As it can be seen from \Fig{fig3}, the $v_{2}\{2k\}~(k=2,3,4)$ coefficients are slightly
correlated to the corresponding initial eccentricities $\varepsilon_{2}\{2k\}$ in all cases.
In this work, we focus on the properties of fluctuations estimated by Pearson coefficient, but not on the response coefficients.

\begin{figure*}[t]
\begin{center}
\includegraphics[width=0.32\textwidth]{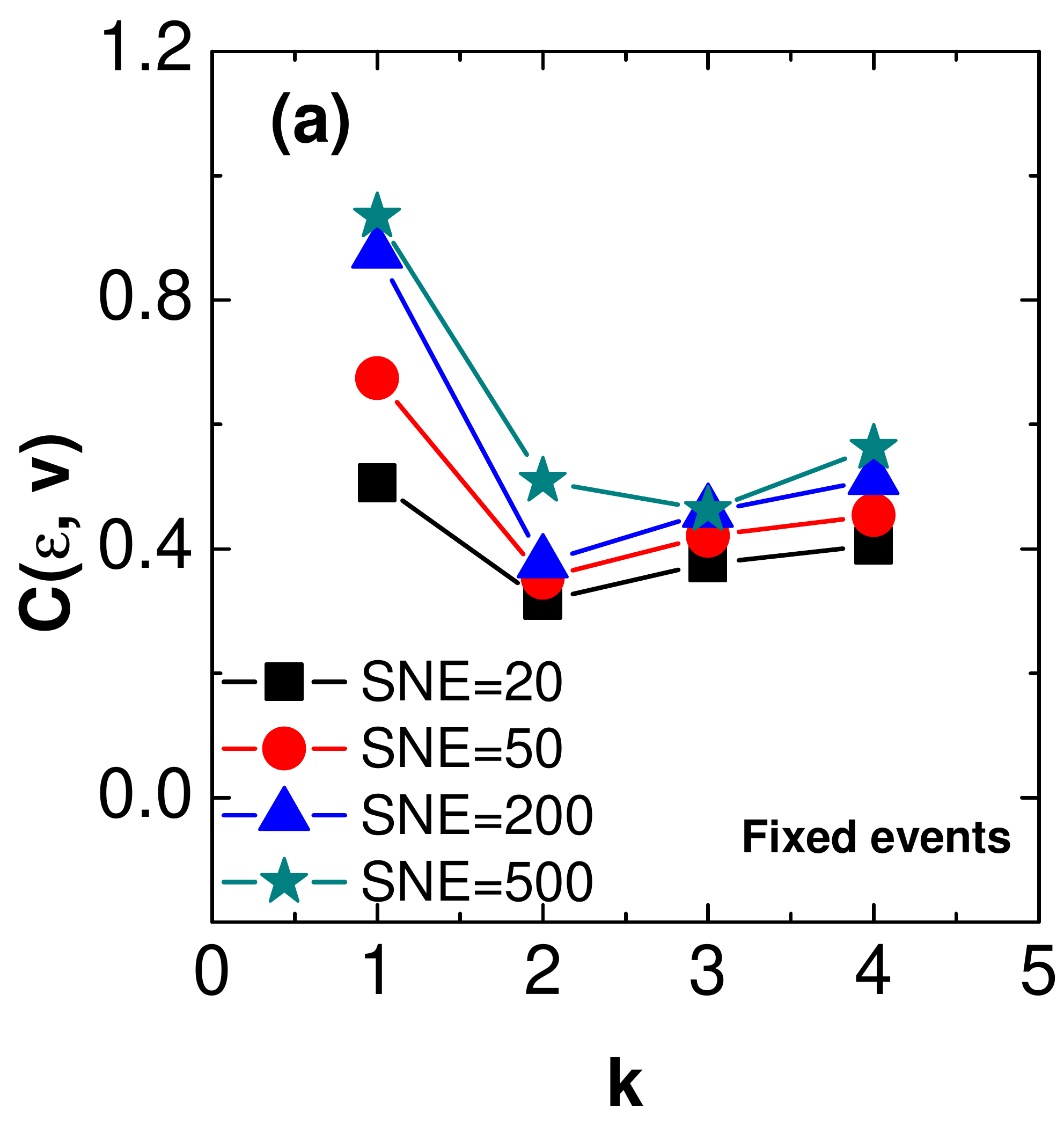}
\includegraphics[width=0.32\textwidth]{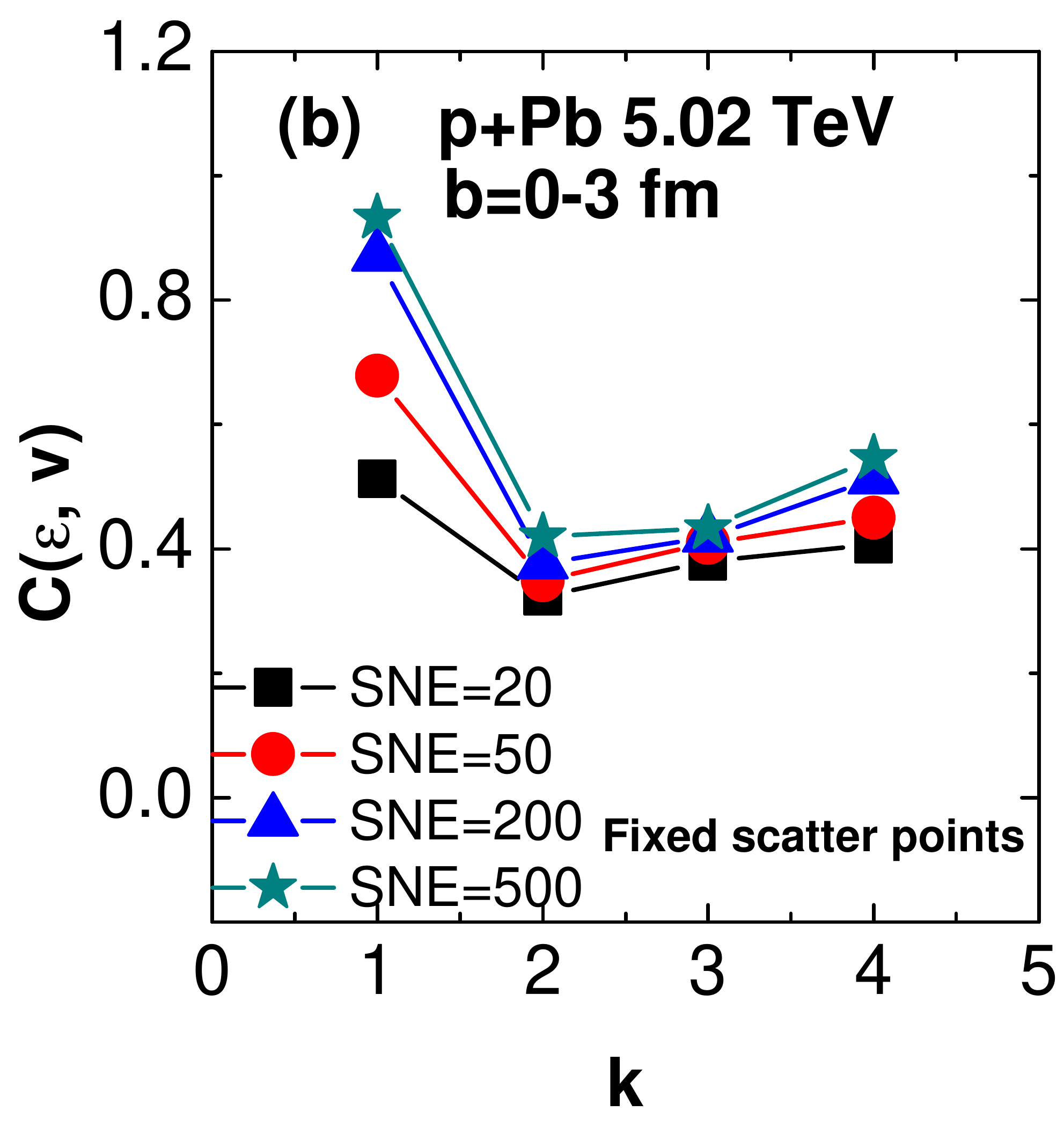}
\includegraphics[width=0.31\textwidth]{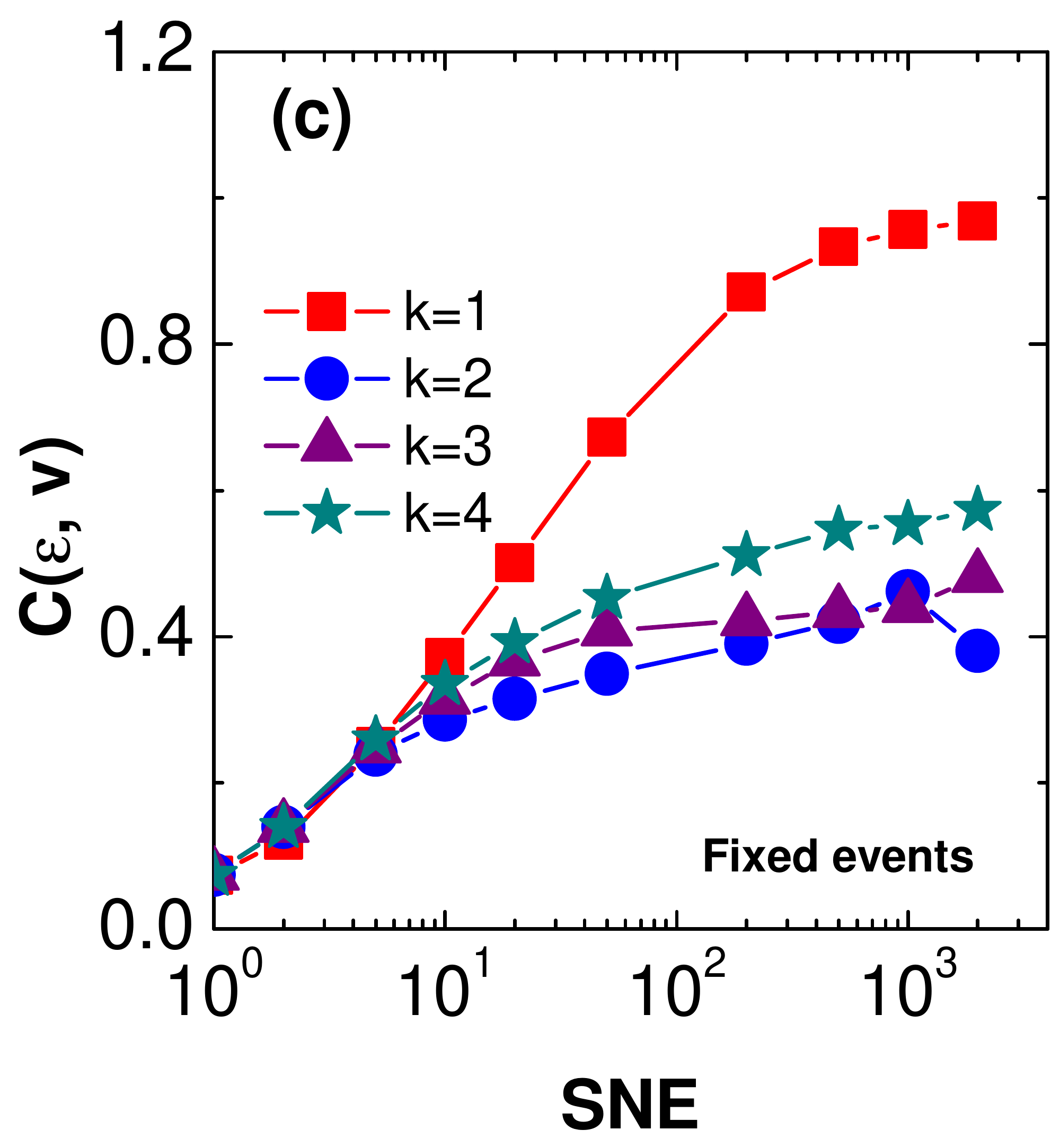}
\caption{(Color online)
Pearson coefficient as a function of the order of cumulants
in p-Pb (b=0-3 fm) collisions. (a) Pearson coefficient at fixed number of
events (from \Fig{fig3}), (b) Pearson coefficient at fixed scatter points (from \Fig{fig4}),
and (c) SNE-dependent Pearson coefficient (at fixed number of events).
}
\label{fig5}
\end{center}
\end{figure*}

Figure~\ref{fig1} and~\ref{fig3} show that the SOE cumulants in
\eq{eq:localcumulants} correspond to a thinner distribution of scattered points
compared to the event-by-event method in \eq{eq:vn} and \eq{eq:en}.
This means that the Pearson coefficient is larger (and correlations stonger)
in the ESC basis rather than in event-by-event basis in this small systems.
For SNE=500, we have $C(\varepsilon, v)\approx 0.560$ for four-particle, six-particle, and eight-particle cumulants, whereas $C(\varepsilon, v)\approx 0.07$ in event-by-event basis.
We show the corresponding Pearson coefficient in \Fig{fig5}. In that case $C(\varepsilon, v)\approx 0.560$, i.e., the harmonics are strongly dependent on the initial states.
For larger SNE number, the shape of the scattered points cloud weakly depends on the order of multi-particle ($k>$1) cumulants.
It has been pointed out that multi-particle cumulants method may suppress fluctuations~\cite{Sirunyan:2019mcs}.
This is confirmed by our analysis, where ESC is also suppresses the fluctuations at different SNE parameters.

As a matter of fact, the cumulants framework may have limitations in describing
$v_{n}$ fluctuations using a small set of cumulants~\cite{Jia:2014efa} due to
the sizeable systematic uncertainties. As a consequence, we evaluate SOE cumulants
to explore the effects of ESC fluctuations for a given impact parameter.
As we mentioned above, fluctuations are suppressed in multi-particle cumulants. The Pearson coefficients of multi-particle
cumulants ($k>$1) are smaller than the Pearson coefficient of two-particle cumulants ($k=$1).
It should be also pointed out that in small systems response may be far from linear,
and our results indicate that current linear response analysis may need to
be carefully re-examined. We are planning to investigate this point elsewhere.

Looking at \Fig{fig3}, one sees that correlations are weaker for low values of the SNE.
In order to better investigate this point, we use a fixed number of scatter points for
different SNE parameters, which are shown in \Fig{fig4}. The total events are set as
5$\times 10^{5}$, 1.25$\times 10^{6}$, 5$\times 10^{6}$and 1.25$\times 10^{7}$, respectively,
which corresponds to SNE=20, 50, 200 and 500, respectively. For these four different SNE
parameters, the number of scatter points is set by consistency.
Rather surprisingly, the distributions in \Fig{fig4} are similar to the distributions
in \Fig{fig3}. This means that the Pearson coefficients do not depend on the sample size
(we show the Pearson coefficients in \Fig{fig5} (a) and (b)).

\begin{figure*}%[tp]
\begin{center}
\includegraphics[width=0.920\textwidth]{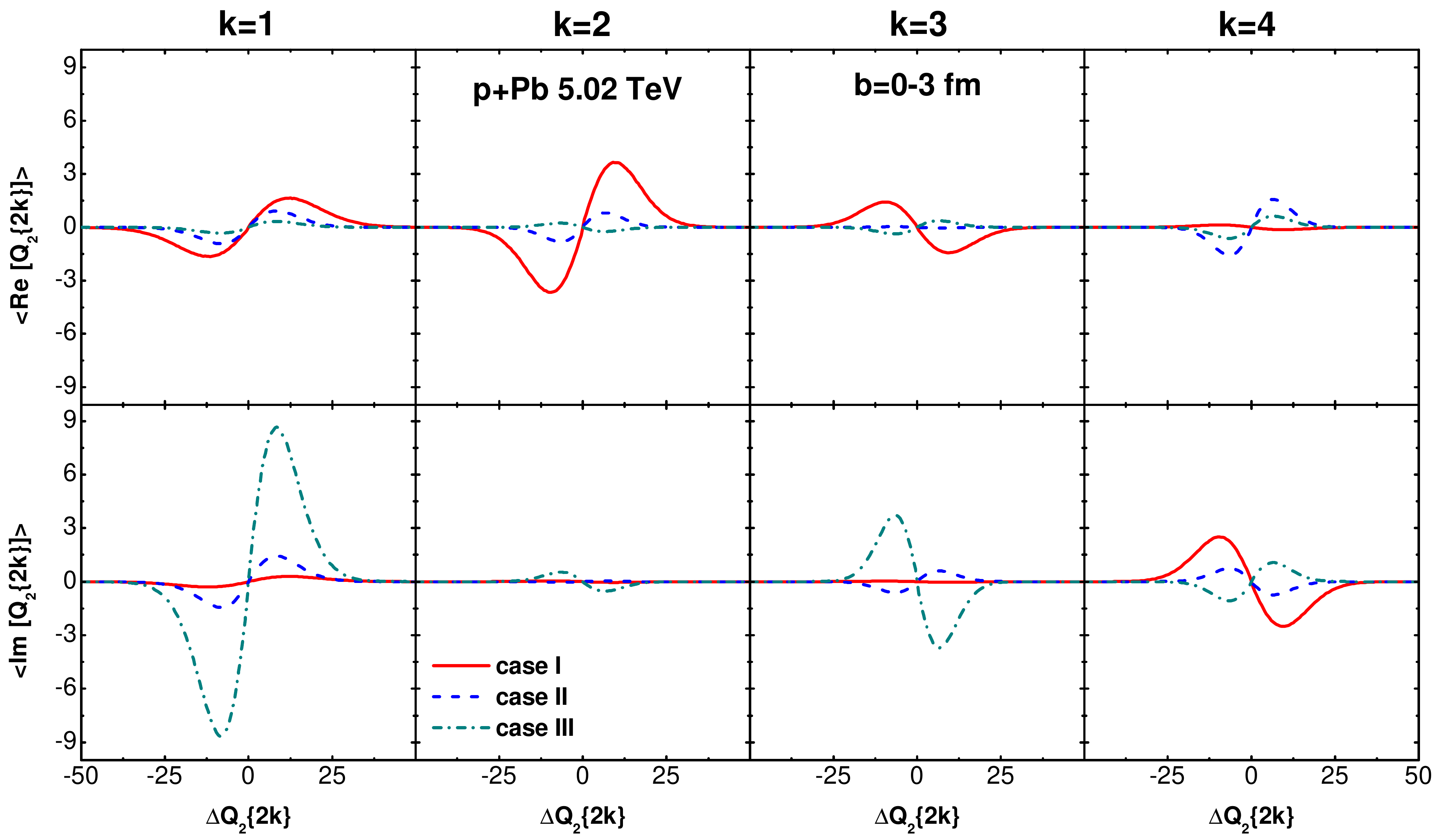}
\caption{(Color online)
Up panels: the averaged distribution of $Re~Q$ (real part of Q-vectors) as functions of Q-vectors difference $\triangle Q_{2}\{2k\}$ in p-Pb (b=0-3 fm) collisions at $\sqrt{s_{NN}}$= 5.02 TeV.
Down panels: similar system with the up panels, but for the averaged distribution of $Im~Q$ (imaginary part of Q-vectors).
}
\label{fig6}
\end{center}
\end{figure*}

In order to evaluate the fluctuations of correlations, we use the Pearson coefficient
$C(\varepsilon, v)$ and assess the linear correlation between the subset
of $v_{2}\{2k\}~(k=1,2,3,4)$ and subset of $\varepsilon_{2}\{2k\}$.
The Pearson coefficients $C(\varepsilon, v)$ are extracted from \Fig{fig3}
and \Fig{fig4} using \eq{eq:pearsonc}.
\Fig{fig5} (a) and (b) shows the Pearson coefficients as functions of the order of multi-particle cumulants in p-Pb (b=0-3 fm) collisions, respectively.
SNE are set to 20, 50, 200 and 500, respectively.
Furthermore, \Fig{fig5} (c) shows the Pearson coefficients as a function
of SNE in p-Pb (b=0-3 fm) collisions.
The Pearson coefficients are significantly-dependent on the SNE number.
They increase with increasing SNE, and achieve their maximum platform
at SNE $\approx$ 1000, as shown in \Fig{fig5} (c).
The Pearson coefficients are weakly-dependent on the order of multi-particle cumulants, except for the two-particle cumulants.
As noted that these two-particle cumulants in the standard framework still retain the effects of unpredicted nonlinear response and fluctuations.
Again, we focus on suppressions of fluctuations with the subset of standard cumulants method.
Even if we increase the SNE, these Pearson coefficients are not normalized (with SNE$\geq$1000),
because: (I) they include also the nonlinear dependence of the final harmonic on the
initial eccentricity (see in \Fig{fig4}), which has not been yet clarified~\cite{Rao:2019bpo,De:2020rli};
(II) the fluctuating final states include contributions not only from the initial fluctuations but also from the dynamical evolution stages~\cite{Wei:2021pot}.
To fully quantify the fluctuations suppression effect caused by the multi-particle cumulants method,
it may need more in-depth research and experimental data.

Here, to prove that the ESC method is not a simple result of events averaged in suppressing fluctuations, we show the results of Q-vector averaged distribution, $\langle Q_{2}\{2k\}\rangle$~($k$=1, 2, 3, 4) in three different cases in~\Fig{fig6}. For~\Fig{fig6}, Case I: SNE=1 (noted as event-by-event simulations) for ESC; Case II: SNE=500 for the subset of cumulants, but without eccentricity re-sorted; Case III: SNE=500 for the subset of cumulants (this ESC method), but with eccentricity re-sorted. The results of up panels are the averaged distribution of Q-vectors (real part) in p-Pb (b=0-3 fm) collisions at $\sqrt{s_ {NN}}$= 5.02 TeV and the results of down panels are the averaged distribution of Q-vectors (image part) in the same system. One can see that, the $\langle Q_{2}\{2k\}\rangle$ as functions of the $Q_{2}$ difference $\triangle Q_{2}\{2k\}=Q_{2}^{i}\{2k\}-\langle Q_{2}\{2k\}\rangle$ ($k$=1, 2, 3, 4,~and $i$ is the $i$-th sample of SOE). The $\triangle Q_{2}\{2k\}$-dependent averaged distribution is inconsistent in these three cases, which shows that the ESC method is different from the conventional subset of cumulants (without concentration resorted) method in suppressing fluctuations.

%=======================================summary=========================================
%\section{Summary and discussions}
\section{Conclusions}
\label{sec:sum}

In conclusion, we have put forward a novel approach to analyze p-Pb (b=0-3 fm) collisions
at $\sqrt{s_{NN}}$= 5.02 TeV, based on ESC response and AMPT model simulations.
We have found that the value of the Pearson coefficients between the subset cumulants of
final harmonics $v_{2}\{2k\}~(k=1,2,3,4)$ and the subset cumulants of initial eccentricity $\varepsilon_{2}\{2k\}$ in the ESC basis is significantly enhanced. These Pearson coefficients
are strongly-dependent on the SNE, and weakly-dependent on the order
of multi-particle cumulants.

In our analysis, the use of ESC method leads to a suppression of the event-by-event fluctuations. Such suppression effects are more significant for
larger SNE, e.g., the fluctuations for SNE$\geq$ 1000 are clearly more suppressed compared
to the case SNE=1. In addition, by analyzing the cumulants for different SNE parameters, we have seen that
the Pearson coefficients are clearly-dependent on the charged multiplicity.

Our results show that ESC analysis provides an enhancement of Pearson coefficients and, in turn,
a mean for studying the fluctuations of fluidlike QGP in heavy-ion collisions.
%
%%%%%%%%%%%%%%%%%%%%%%%

\section*{Acknowledgements}
D.-X. W. has been supported by the National Natural Science Foundation of China Grant No.~12105057, the Youth Program of Natural Science Foundation of Guangxi (China) Grant No.~2019GXNSFBA245080, the Special fund for talents of Guangxi (China) Grant No.~AD19245157, and the Doctor Startup Foundation of Guangxi University of Science and Technology Grant No.~19Z19. L.-J. Z. has been supported by the National Natural Science Foundation of China Grant No.~11865005 and the Natural Science Foundation of Guangxi (China) Grant No.~2018GXNSFAA281024.

%\bibliography{factor_refs}

\end{document}